\documentclass[aps,pra,superscriptaddress,twocolumn,showpacs,10pt]{revtex4-1} 

\usepackage{amsmath,amssymb,graphicx}
\usepackage{todonotes}
\usepackage{subfigure}
\usepackage[export]{adjustbox}
\usepackage{array}
\usepackage[english]{babel}

\begin{document}
\title{Femtosecond Self-Reconfiguration of Laser-Induced Plasma Patterns in Dielectrics}
\author{Jean-Luc D\'eziel}
\affiliation{D\'epartement de physique, de g\'enie physique et d'optique, Universit\'e Laval, Qu\'ebec G1V 0A6, Canada}
\author{Louis J. Dub\'e}
\email{Louis.Dube@phy.ulaval.ca}
\affiliation{D\'epartement de physique, de g\'enie physique et d'optique, Universit\'e Laval, Qu\'ebec G1V 0A6, Canada}
\author{Sandra H. Messaddeq}
\author{Youn\`es Messaddeq}
\affiliation{Centre d'Optique, Photonique et Laser, Universit\'e Laval, Qu\'ebec G1V 0A6, Canada}
\author{Charles Varin}
\affiliation{D\'epartement de physique, de g\'enie physique et d'optique, Universit\'e Laval, Qu\'ebec G1V 0A6, Canada}
\affiliation{D\'epartement de physique, Universit\'e d'Ottawa, Ontario K1N 6N5, Canada}
\affiliation{C\'egep de l'Outaouais, Gatineau, Qu\'ebec J8Y 6M4, Canada}

\begin{abstract}
Laser-induced modification of transparent solids by intense femtosecond laser pulses allows fast integration of nanophotonic and nanofluidic devices with controlled optical properties. Experimental observations suggest that the local and dynamic nature of the interactions between light and the transient plasma plays an important role during fabrication. Current analytical models neglect these aspects and offer limited coverage of nanograting formation on dielectric surfaces. In this paper, we present a self-consistent dynamic treatment of the plasma build-up and its interaction with light within a three-dimensional electromagnetic framework. The main finding of this work is that local light-plasma interactions are responsible for the reorientation of laser-induced periodic plasma patterns with respect to the incident light polarization, when a certain energy density threshold is reached. Plasma reconfiguration occurs within a single laser pulse, on a femtosecond time scale. Moreover, we show that the reconfigured sub-wavelength plasma structures actually grow into the bulk of the sample, which agrees with the experimental observations of self-organized volume nanogratings. We find that mode coupling of the incident and transversely scattered light with the periodic plasma structures is sufficient to initiate the growth and self-organization of the pattern inside the medium with a characteristic half-wavelength periodicity.
\end{abstract}


\maketitle

\section{Introduction}

Laser micro-machining of transparent material by femtosecond pulses promises fast integration of 2D and 3D nanophotonic and nanofluidic devices with tailored properties \cite{tan2016,gattass2008,taylor2008,qiu2008}. Exposition of solid materials to intense laser radiation typically gives rise to complex light-matter interaction processes like high harmonic generation~\cite{vampa2015}, ionization avalanche breakdown \cite{stuart1995}, and ultrafast plasma dynamics \cite{varin2012}. Such nonlinear light-matter interaction processes are often associated with instabilities that lead to the formation of periodic surface and bulk patterns \cite{bhardwaj2006,bonse2017}. Thus far, static surface mode analysis has successfully explained the origin and characteristics of periodic patterns formed on laser-processed metals and semi-conductors \cite{bonse2017,bonse2012}. However, better understanding of femtosecond laser-plasma interaction dynamics is necessary to explain nanograting formation in dielectrics, where predictions from current theoretical models are inconsistent with respect to experimental observations on several important aspects.

Extensive experimental and theoretical studies have examined a number of laser-induced nanostructures that arise from the interaction between intense femtosecond laser pulses and materials. In particular, laser-induced periodic surface structures (LIPSS) have been studied with a model proposed by Sipe \cite{sipe1983} and explained as the result of interference patterns between the incident light and static surface modes triggered by surface inhomogeneities. A large number of experimentally observed LIPSS on lossy materials are correctly described within this framework~\cite{bonse2012}. However, it lacks proper account of the plasma dynamics that characterizes the interaction of intense femtosecond pulses with dielectrics. To account for the plasma formation, an extended Sipe theory has been explored \cite{bonse2009,hohm2012}, but in these studies the plasma is assumed to be static and homogeneous, whereas during the ionization of transparent materials the carrier density changes rapidly in time and can exhibit significant spatial variations. For a rigorous analysis, transient optical properties, referred to as \emph{intrapulse feedback}, have to be taken into account to allow for self-interaction and therefore, self-organization of plasma patterns while the plasma is being formed. 

Light-matter interaction feedback can also occur between subsequent laser pulses, referred to as \emph{interpulse feedback} \cite{Skolski2014}. The target can be structurally modified by each laser pulse by mechanisms like ion removal, chemical bonding reconfiguration, melting, hydrodynamic flow, crystallization, amorphization, and so on \cite{gamaly2002,costache2004}. The next laser pulses then interact with these permanent modifications and interfere constructively or destructively with them \cite{deziel2016}. Interpulse mechanisms result from the inhomogeneous energy distribution after irradiation, itself determined by intrapulse dynamics. Therefore, laser-plasma interplay on the single-pulse scale has to be fully understood first, before attempting a complete description of nanostructures fabricated with multiple laser pulses.

In the accepted interpretation, single-pulse energy deposition in metals or ionized materials occurs in two steps, where energy is first transferred to free or nearly-free electrons and then from the free electrons to the lattice (see, e.g.,~\cite{gamaly2013,tan2016}). Accurate knowledge of the plasma distribution and dynamics is therefore important to understand the entire light-matter process. This is particularly true in dielectrics, where the electron density changes in time and is spatially inhomogeneous due to the local nature of the underlying ionization processes. 

Studies that accounted for the time and space dependence of the plasma density reported self-organization of plasma patterns in the bulk of dielectric materials \cite{bhardwaj2006,bulgakova2013}. However, computer simulations with the proposed theoretical \emph{nanoplasmonic} model predict plasma growth in the direction opposite to the laser propagation, with periodic width oscillations along the same axis \cite{buschlinger2014,rudenko2016}. In contrast, experiments show that structures grow towards the laser propagation direction with a constant width \cite{taylor2008}.

In this work, we present a self-consistent dynamic treatment of the interaction of a femtosecond laser pulse with a dielectric medium. The spatio-temporal dependence of the light-induced near-surface plasma density build-up and patterning was studied within a three-dimensional electromagnetic framework. An important result of this research is the observation of an ultrafast reconfiguration of the plasma distribution, when the conditions for self-organization are met. Moreover, through an extended analysis, we obtain a bulk growth of the plasma patterns in the same direction as the laser propagation with constant width, in agreement with experimental observations of volume nanogratings \cite{taylor2008}. This correct plasma growth is observed when intrapulse self-organization is dominated by transverse scattering against transient plasma structures. In comparison, in the current nanoplasmonic model, longitudinal scattering is dominant, which seems to lead to incorrect predictions. Our work confirms that a three-dimensional dynamical treatment of the plasma density is crucial to extend the LIPSS theory to transparent medium.

The paper is divided as follows. First in Sec.~\ref{sec:model}, we present the details of the numerical model and provide specific simulation parameters. Then in Sec.~\ref{sec:patterns}, we show how early-interaction plasma patterns aligned along the laser pulse electric field reorient themselves within only a few femtoseconds, when a certain fluence threshold is reached. In Sec.~\ref{sec:robustness}, we discuss the robustness of the self-reorganisation phenomena with respect to various parameters and the simple ionization model introduced in Sec.~\ref{sec:model}. We further discuss in Sec.~\ref{sec:reflections} the origin of the reorganized plasma patterns and show how they grow in the bulk. In Sec.~\ref{sec:origins}, we present a spectral analysis and identify four distinct categories of patterns, some of which can be explained with Sipe theory, whereas others are characteristic of the electromadynamic model we present. Finally, conclusions are given in Sec.~\ref{sec:conclusions}.

\section{Numerical model\label{sec:model}}

\subsection{Electromagnetic fields and currents}

For this work, we build a phenomenological model of the interaction of an intense femtosecond laser pulse with a dielectric material. We used the finite-difference time-domain (FDTD) method \cite{taflove2005} to solve Maxwell equations
\begin{align}
&\nabla \times \vec{E} = -\mu_0\frac{\partial \vec{H}}{\partial t}, \label{max1}\\
&\nabla \times \vec{H} = \epsilon_0\frac{\partial \vec{E}}{\partial t} +\vec{J},\label{max2}
\end{align}
$\vec{E}$, $\vec{H}$ and $\vec{J}$ are respectively the electric field, the magnetic field and the total electric current density. Self-consistent material effects were included via three contributing currents. (i) A current density from bound electrons
\begin{align}
\vec{J}_b=\epsilon_0 \frac{\partial}{\partial t} \left(\chi^{(1)}\vec{E}+\chi^{(3)}|\vec{E}|^2 \vec{E} \right) \label{Jb}
\end{align}
accounts for first and third order susceptibilities (see~\cite{varin2015,varin2018} for details). (ii)~The free-carrier current density
\begin{align}
\frac{\partial \vec{J}_f}{\partial t} = -\gamma\vec{J}_f + \epsilon_0 \omega_p^{2}\vec{E}
\end{align} 
accounts for the plasma optical response with $\omega_p^2 = q^{2}\rho/(\epsilon_0 m_e^*)$, where $m_e^{*}$ is the effective mass of the electron and $\rho$ is the time and space dependent plasma density. The time integration of $\vec{J}_f$ is explicitly implemented in our finite difference calculation with
\begin{align}
\vec{J}_f^{m+1} = \frac{g_1}{g_2}\vec{J}_f^{m} + \frac{\epsilon_0\omega_p^{2}\delta_t}{g_2}\vec{E}^{m},
\end{align}
where $m$ is the time step index, $g_1=(1-\gamma\delta_t/2)$, $g_2=(1+\gamma\delta_t/2)$ and $\delta_t$ is the temporal discretization parameter. (iii) The third current accounts for field energy losses due to field ionization, calculated with 
\begin{align}
\vec{J}_K = \frac{\mathcal{E}_g \nu_K (\rho_{\mathrm{mol}}-\rho)}{|\vec{E}|^{2}}\vec{E}. \label{JK}
\end{align}
Where $\mathcal{E}_g$ is the bandgap of the transparent material, $\nu_K$ is the field ionization rate and $\rho_{\mathrm{mol}}$ is the molecular density. The total current density is then calculated with $\vec{J}=\vec{J}_b+\vec{J}_f+\vec{J}_K$. 

\subsection{Ionization mechanisms}

Hereafter, we use a dimensionless plasma density by normalizing with respect to the molecular density $\bar{\rho} = \rho/\rho_{\mathrm{mol}}$. Considering only the first ionization state, the plasma formation is calculated with
\begin{align}
\frac{\partial \bar{\rho}}{\partial t}= \left(\nu_K+\nu_C \right) (1-\bar{\rho}), \label{sre}
\end{align}
where the factor $(1-\bar{\rho})$ ensures that saturation is reached when all molecules are ionized. The field ionization rate $\nu_K$ accounts for multiphoton and tunnel ionization mechanisms and is calculated with the Keldysh model adapted for transparent material \cite{keldysh1965}. In our simulations, the Keldysh parameter $\gamma_K = \omega \sqrt{m^{*}\mathcal{E}_g}/(q|\vec{E}|)$ extends from values much larger than 1 and $\gamma_K \sim 0.4$, which respectively belong within the multiphoton and the tunnel ionization regimes. This means that both field ionization mechanisms are relevant and must be accounted for.

A simple approach to calculate the collisional ionization rate $\nu_C$, the Single Rate Equation (SRE) \cite{stuart1995}, considers that every free electron can contribute to $\nu_C$ proportionally with the local light intensity $I(t)=c\epsilon_0n |\vec{E}(t)|^{2}$, thus, $\nu_C = \alpha \bar{\rho} I(t)/2$ with the collisional cross-section $\alpha$ \footnote{The factor $1/2$ is added in the definition of $\nu_C$ for consistency with \cite{stuart1995} in which the intensity is period averaged, therefore, half the instantaneous intensity we use.}. A more accurate model, the Multiple Rate Equations (MRE) \cite{rethfeld2004}, accounts for the transient energy distribution of the electrons in the conduction band (CB) and the impact rate between the most energetic free electrons and neutral molecules that leads to collisional ionization. The main difference when using MRE over SRE is the inhibition of collisional ionization for very short pulse durations, for which free electrons do not have the time to build up enough energy. Practically, this results in an implicit dependency between the collisional cross-section $\alpha$ and the pulse duration $\tau$. Because we do not study multiple pulse durations, the use of MRE is not required here. We stress that values of $\alpha$ used in the literature vary
considerably. This particular aspect is assessed in Sec. \ref{sec:robustness}.

\subsection{Material, geometry and laser source}

In our simulations, material parameters are chosen to reproduce the optical properties of fused silica; first and third order susceptibilities $\chi^{(1)}=1.1025$ (refractive index $n=1.45$), $\chi^{(3)}=2\cdot 10^{-18}$ cm$^2/$V$^2$, plasma damping $\gamma=10^{15}$ s$^{-1}$, effective mass of the electron $m_e^*=0.8m_e$, molecular density $\rho_{\mathrm{mol}}=2.2\cdot 10^{22}$ cm$^{-3}$ and band-gap $\mathcal{E}_g=9$ eV \cite{christensen2009,jia2017}. For the collisional cross-section, we first use the highest value found in the literature, $\alpha = 10$ cm$^{2}$/J from \cite{stuart1995}, and then compare with the results obtained when using $\alpha = 0$ cm$^{2}$/J.

\begin{figure}
\centering
\includegraphics[width=1.0\columnwidth]{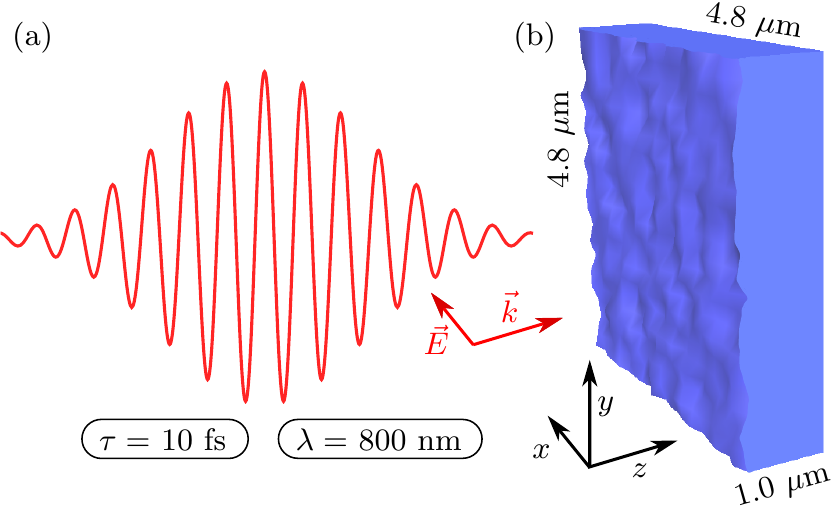}
\caption{(Color online) Schematic illustration of the simulations. In (a), the laser pulse is linearly polarized along the $x$-axis and propagates towards the $+z$-axis. It hits, with normal incidence, a fused silica sample with a rough surface, shown in (b).}\label{figDomain}
\end{figure}

The geometry of the simulations is schematized in Fig. \ref{figDomain}. The target is a square of side 4.8 $\mu$m with a thickness of 1 $\mu$m. The laser source, expressed as
\begin{align}
\vec{E}(t)|_{\mathrm{source}} = E_0 e^{-(t/\tau)^{2}} \sin(\omega t) \hat{e}_x, 
\end{align}
is a plane wave linearly polarized along the $x$-axis propagating along the $+z$-axis with a wavelength of $\lambda = 2\pi c/\omega = 800$ nm. Laser pulses have a gaussian time envelope of duration $\tau=10$ fs. The laser fluence, defined as
\begin{align}
F = \int_{-\infty}^{\infty} I(t)\mathrm{d}t = \sqrt{\frac{\pi}{2}} \frac{c\epsilon_0\tau}{2}E_0^{2} ,
\end{align}
is the energy per unit area. We ran simulations using fluence values from 0.7 to 5 J/cm$^{2}$.

Spatial discretization parameters are $\delta_{x,y}=20$ nm and $\delta_z=2$ nm. Surface roughness is mimicked by randomly adding 0 to 20 nm of material over every discrete coordinate of the surface, for an average height of 10 nm. The temporal discretization $\delta_t=6.33\cdot 10^{-3}$ fs is chosen to be below the instability regime of the FDTD algorithm. The time domain extends from $t=-1.5\tau$ to $t=+1.5\tau$. To run the simulations, we used the open software EPOCH \cite{epoch}, extended to include the Eqs. \eqref{Jb} to \eqref{sre}. 

\section{Self-reconfiguration of plasma patterns\label{sec:patterns}}

\begin{figure*}
\centering
\includegraphics[width=1.0\textwidth]{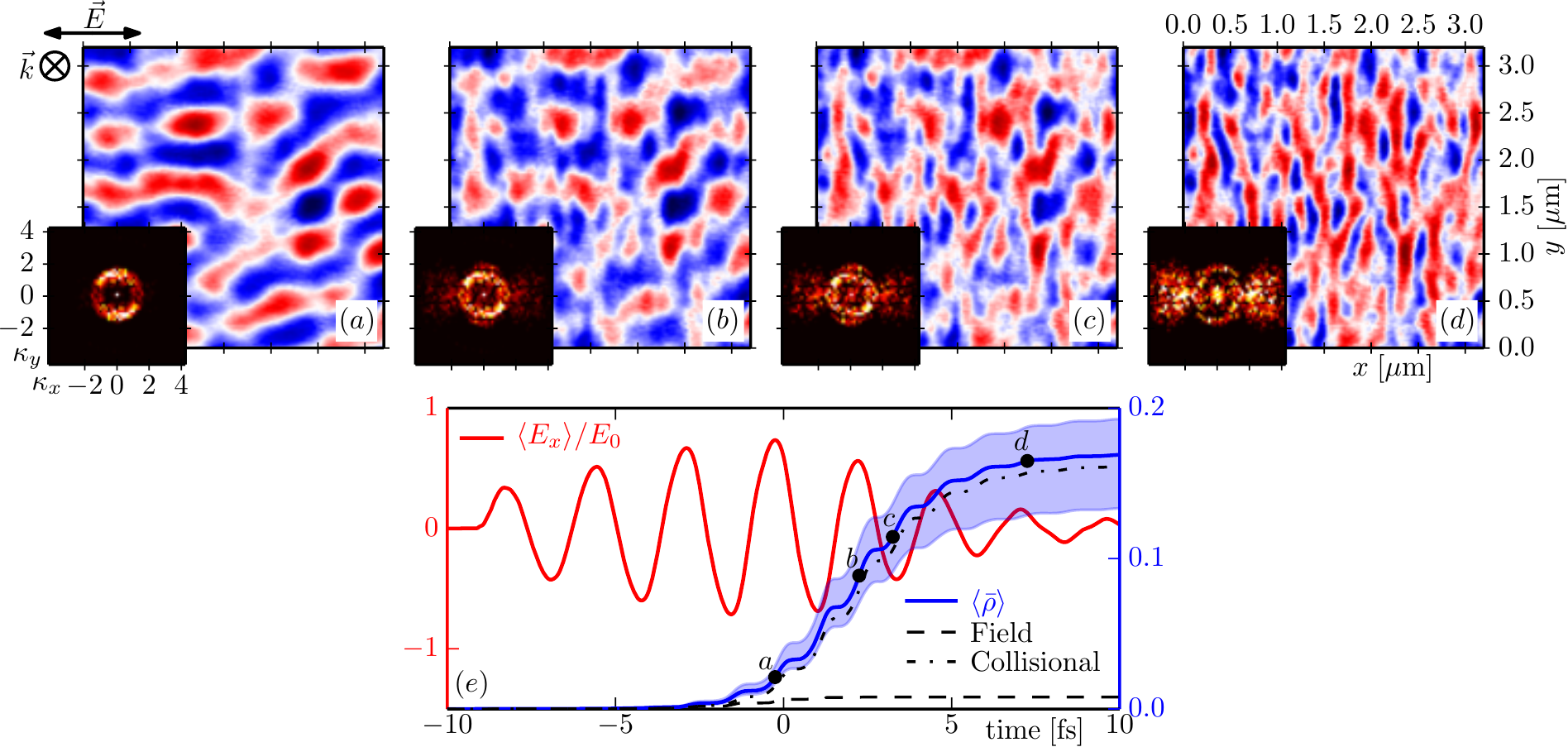}
\caption{(Color online) Top row (a) to (d) : Time evolution of the variations in plasma density $\rho(x,y,z)$ (from blue minima to red maxima) at fixed depth $z=140$ nm with $\tau=10$ fs, $\alpha=10$ cm$^{2}$/J and $F=2.12$ J/cm$^2>F_{\mathrm{th}}$ showing the structural transition from $\Lambda_\parallel\sim\lambda/n$ to $\Lambda_\perp\sim\lambda/2n$. Overlayed subfigures show the 2D Fourier transforms of the corresponding figure. Wavenumbers in the Fourier space $\vec{\kappa}$ are all in units of $\lambda^{-1}$. (e): $E_x$ (normalized with respect to $E_0$) and $\bar{\rho}$ averaged over the $(x,y)$ plane at $z=140$ nm. The blue shade indicates the amplitude of the variations of $\bar{\rho}$, magnified 10 times to make them more apparent on this scale. The black dots indicate the times of the snapshots (a) to (d).}\label{figTimeEvolution}
\end{figure*}

The most striking result obtained with the electrodynamic model presented in Sec.~\ref{sec:model} is the observation of an ultrafast reconfiguration of the plasma pattern occurring during ionization (see Fig.~\ref{figTimeEvolution}). The plasma structures initially oriented parallel ($\parallel$) to the laser polarization and with a periodicity $\Lambda_\parallel\sim\lambda/n$ change to orthogonal ($\perp$) structures with $\Lambda_\perp\sim\lambda/2n$ within a few optical cycles, where $n$ is the refractive index. This transition occurs approximatively when the laser fluence is high enough for the plasma to reach critical density $\bar{\rho}_c = \epsilon_0 m^{*} \omega^{2}/q^{2} \sim 6.3$\% (corresponding to $\omega_p = \omega$), defining as such a fluence threshold $F_\mathrm{th}$. These two patterns are commonly observed in experiments with dielectrics and the theory has never been able to account for both in a single framework. Whereas LIPSS theory predicts the former pattern ($\Lambda_\parallel\sim\lambda/n$), the nanoplasmonic model predicts the latter ($\Lambda_\perp\sim\lambda/2n$).

In the early stage of the interaction [see Fig. \ref{figTimeEvolution}(a)], the plasma density is too low for the transient optical properties to cause noticeable feedback and induce self-reorganisation. In this regime, Sipe's static surface mode analysis \cite{sipe1983} describes correctly the early plasma growth and emergence of periodic plasma patterns with $\Lambda_\parallel\sim\lambda/n$. These structures result from the interference between the incident light and the scattered radiation off the surface inhomogeneities. However, as the average carrier density increases and intrapulse feedback gets stronger, these characteristic patterns gradually disappear [see Fig. \ref{figTimeEvolution}(b)-(d)], ultimately to leave the stage to patterns perpendicular to the laser polarization with $\Lambda_\perp\sim\lambda/2n$. This peculiar transition from one type of pattern to the other is a direct consequence of the intrapulse feedback allowed by our model. The origin and mechanisms behind this ultrafast reorganisation of the plasma is further discussed in section \ref{sec:reflections}.

\subsection{Robustness\label{sec:robustness}}

The results presented in this work are representative of a broad parameter space. With wavelengths between 400 and 1200 nm, pulse durations between 5 and 250 fs, plasma damping between 10$^{14}$ and 10$^{16}$ s$^{-1}$ and effective mass of the electron between 0.5$m_e$ and 1.0$m_e$, we find that self-reconfiguration consistently occurs for $F>F_{\mathrm{th}}$. These parameter variations can indeed shift the absolute value of $F_{\mathrm{th}}$, but this does not affect the occurrence of the events around the threshold, wherever it may be.

\begin{figure}
\centering
\includegraphics[width=1.0\columnwidth]{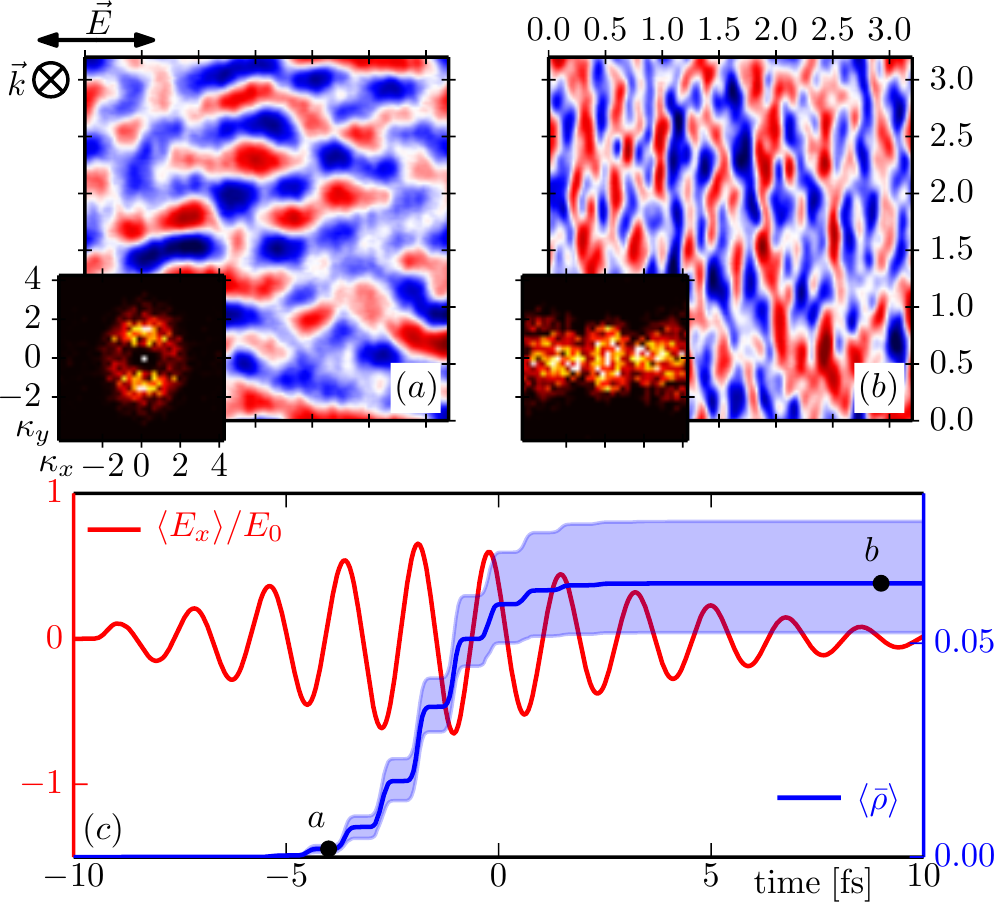}
\caption{(Color online) Variations in plasma density $\rho(x,y,z)$ (from blue minima to red maxima) at fixed depth $z=140$ nm with $\tau=10$ fs and $F=4.95$ J/cm$^2$ showing the structural transition from (a) $\Lambda_\parallel\sim\lambda/n$ to (b) $\Lambda_\perp\sim\lambda/2n$. The collisional cross-section $\alpha$ is set to 0, so only field ionization contributes for plasma formation. (c): $E_x$ (normalized with respect to $E_0$) and $\bar{\rho}$ averaged over the $(x,y)$ plane at $z=140$ nm. The blue shade indicates the amplitude of the variations of $\bar{\rho}$, magnified 10 times to make them more apparent on this scale. The black dots indicate the times of the snapshots (a) and (b).}\label{figNocol}
\end{figure}

As shown in Fig. \ref{figTimeEvolution}(e), most of the plasma is generated through collisional ionization, which might be questionable for such a short pulse duration $\tau$ \cite{rethfeld2006}. However, we find that the nature of the dominant ionization mechanism does not affect significantly the self-reconfiguration process as the same phenomena occurs at $F>F_{\mathrm{th}}$ after setting $\alpha = 0$ cm$^{2}$/J. The same structures are indeed present in Fig. \ref{figNocol} where only field ionization was possible. The value $\alpha=10$ cm$^{2}$/J is therefore used in the following sections.

As for field ionization, more advanced descriptions that account for sub-cycle effects \cite{yudin2001,zhokhov2014} that are absent from the Keldysh formalism are under studies. However, the fact that the self-reconfiguration is consistently occurring, even when driven by vastly different ionization mechanisms (field or collisional ionization), suggests that it has a weak dependence upon the specific description of plasma formation. Moreover, the plasma structures obtained in our simulations agree with experimental observations and results from previous models (in their respective limits), which also suggests that sub-cycle ionization effects should not have a significant impact.

\subsection{The role of reflections off the inhomogeneous plasma density\label{sec:reflections}} 

The electrostatic, nanoplasmonic model currently used to explain the formation of periodic patterns in the bulk relies on an initial distribution of nanovoids to initiate the growth of self-organized plasma patterns with $\Lambda_\perp\sim\lambda/2n$ \cite{bhardwaj2006}. Recent simulations \cite{buschlinger2014,rudenko2016} have shown that around these nanovoids spherical nanoplasmas initially grow, causing strong optical reflections and the formation of standing waves along the longitudinal direction. These standing waves are responsible for further plasma growth in the direction opposite to the laser propagation, in contradiction with experiments \cite{taylor2008}.

\begin{figure}
\centering
\includegraphics[width=1.0\columnwidth]{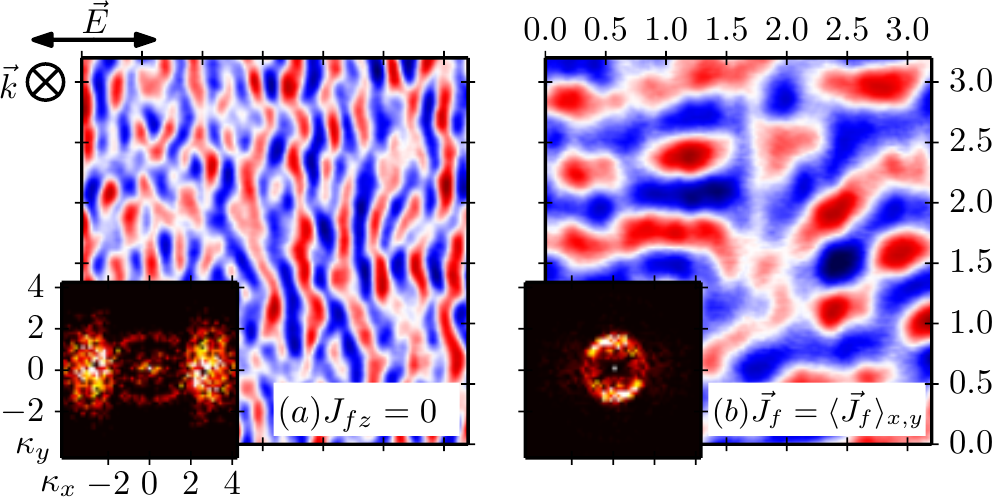}
\caption{(Color online) Final variations in plasma density $\rho(x,y,z)$ (from blue minima to red maxima) when reflections are turned off along (a) the $z-$axis or (b) $x-$ and $y-$axes. All parameters are identical to the ones used in Fig. \ref{figTimeEvolution}.} \label{figReflections}
\end{figure}

To test the role of longitudinal reflections in our simulations, we have excluded the $z$ component of $\vec{J}_f$ in the full simulation and obtained Fig.~\ref{figReflections}(a), where the self-organized and reconfigured plasma structures of Fig.~\ref{figTimeEvolution}(d) are still present. This means that self-organization can be achieved \emph{without} the contribution from longitudinal reflections, previously thought to be necessary for the formation of bulk patterns \cite{rudenko2016}. In contrast, transverse reflections along both orthogonal directions $x$ and $y$ can also be turned off by averaging the free currents $\vec{J}_f(x,y,z)$ over the $(x,y)$ plane, at every time step of the simulation. The corresponding final plasma distribution is shown in Fig. \ref{figReflections}(b).  It is almost identical to Fig.~\ref{figTimeEvolution}(a), which indicates that without transverse reflections, pattern reconfiguration does not occur.

Reflections against local maxima in the plasma density scatter light in the transverse plane. This scattered light couples to the natural modes associated with the plasma density patterns. The interplay between this transversely scattered light, the plasma density, and the incident light may be additive (positive feedback) or subtractive (negative feedback), which leads to the growth or inhibition of periodic structures in both $x$ and $y$ directions. Therefore, the ultrafast pattern reorganization shown in Fig. \ref{figTimeEvolution} can be explained by a metallic waveguide approach, already believed to be linked to self-organized plasma patterns (see, e.g.,~\cite{rajeev2007}).

\begin{figure}
\centering
\includegraphics[width=1.0\columnwidth]{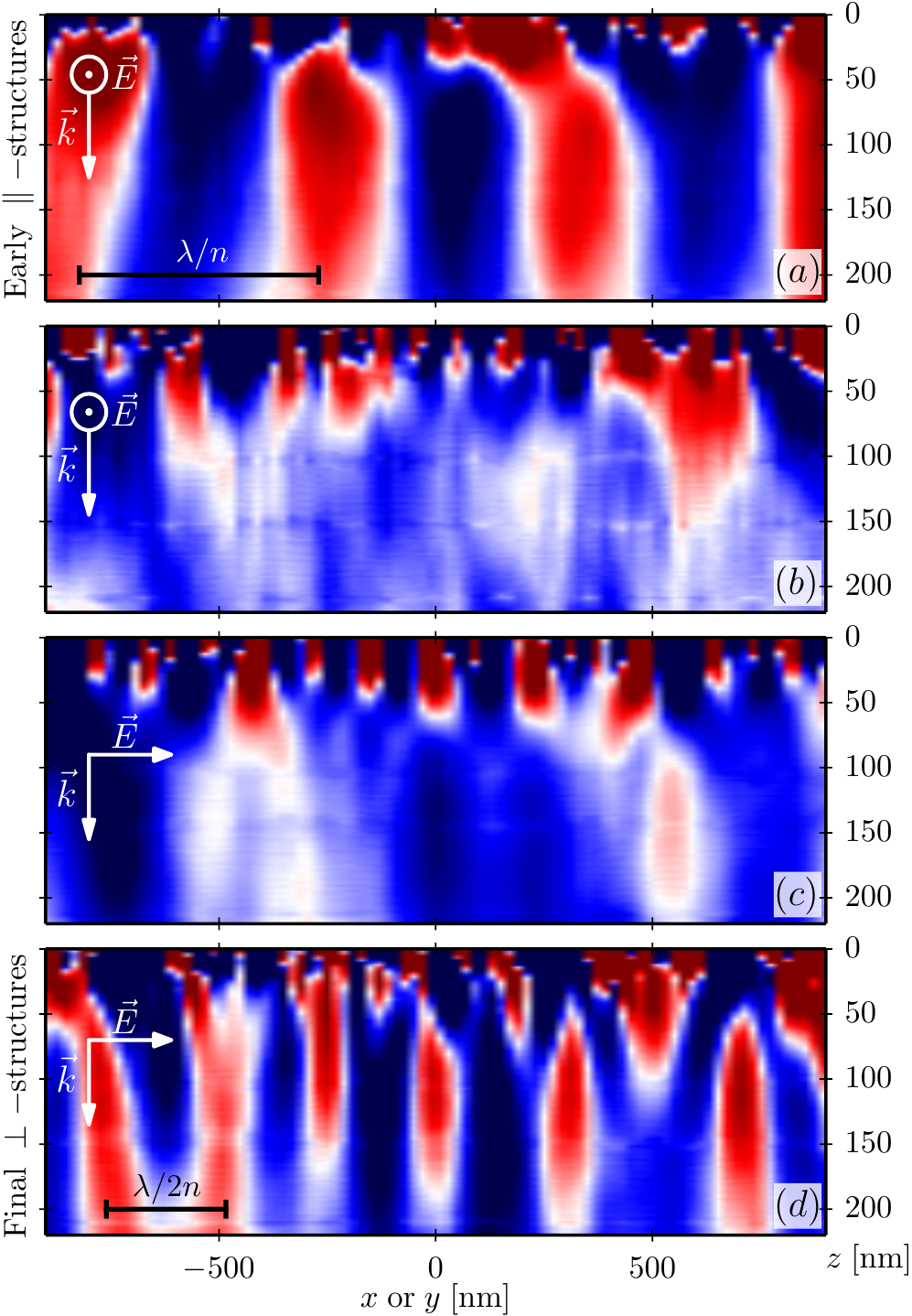}
\caption{(Color online) Transverse views of the relative variations in the plasma density (from blue minima to red maxima) generated with pulse parameters $\tau=10$ fs and $F=2.12$ J/cm$^2$. First, a cut along the $y$-axis in the early $\parallel$-structures (a) and in the final flat distribution (b) caused by negative feedback with the propagated field. Second, a cut along the $x$-axis in the initially flat plasma distribution (c) and in the final $\perp$-structures (d) amplified by positive feedback. }\label{figTrans}
\end{figure}

A consequence of the dynamic laser-plasma interplay just described is the self-inhibition, during ionization avalanche, of the early $\parallel$-structures shown in Figs. \ref{figTimeEvolution}(a) and~\ref{figTrans}(a). We recall that this type of structures, whose origin is explained by Sipe \cite{sipe1983}, appears when the plasma density is low and intrapulse interaction feedback is negligible. However, our simulations show that these structures are dynamically suppressed when the plasma approaches critical density. Our interpretation is the following. (i) The local maxima of the early $\parallel$-structures effectively serve as waveguide boundaries within which the lowest-order transverse electric (TE$_{1}$) resonant modes develop. (ii)~The field distribution of these TE$_{1}$ modes is maximum halfway between the walls, where ionization is locally enhanced. The early plasma density distribution then becomes anti-correlated with ionization, i.e., a negative feedback loop drives the distribution towards a nearly-flat equilibrium [see Figs.~\ref{figTimeEvolution}(b,c) and \ref{figTrans}(b) where the $\parallel$-structures have essentially vanished].

In the orthogonal orientation, the plasma distribution is initially flat in the bulk, as shown in Fig. \ref{figTrans}(c). However, this equilibrium becomes unstable when feedback starts to manifest. As soon as the symmetry of the incident plane wave breaks, some energy falls into the lowest-order non-uniform transverse magnetic (TM$_1$) resonant mode. Since  the TM$_1$ mode has antinodes located at the maxima of the local plasma density, a positive feedback loop is thereby created, amplifying $\perp$-structures [see Fig. \ref{figTrans}(d)].

In the present simulations, the symmetry is broken when light interacts with the surface roughness of the sample. In real bulk interactions, volume defects could also break the symmetry of the incident plane wave to deviate from the initial and unstable flat plasma distribution. However, volume defects can not be responsible for most of the initial plasma formation, as in the nanoplasmonic model, without causing the formation of a strong standing wave along the propagation axis.

In summary, the early plasma structures grow without feedback and with a pattern correctly described by the Sipe theory for transparent medium, that is $\Lambda_\parallel\sim\lambda/n$. When the plasma approaches the critical density, transverse reflections gain in importance and drive the formation of transverse resonant light modes. Negative feedback between $\parallel$-structures and a TE mode essentially cancels the early plasma patterns. At the same time, positive feedback between $\perp$-structures and a TM mode amplifies structural growth in a direction perpendicular to the original orientation [see, again, Fig \ref{figTimeEvolution}(a)-(d)].

\subsection{The characteristics of the plasma patterns}\label{sec:origins} 

\begin{figure}[h!]
\centering
\includegraphics[width=1.0\columnwidth]{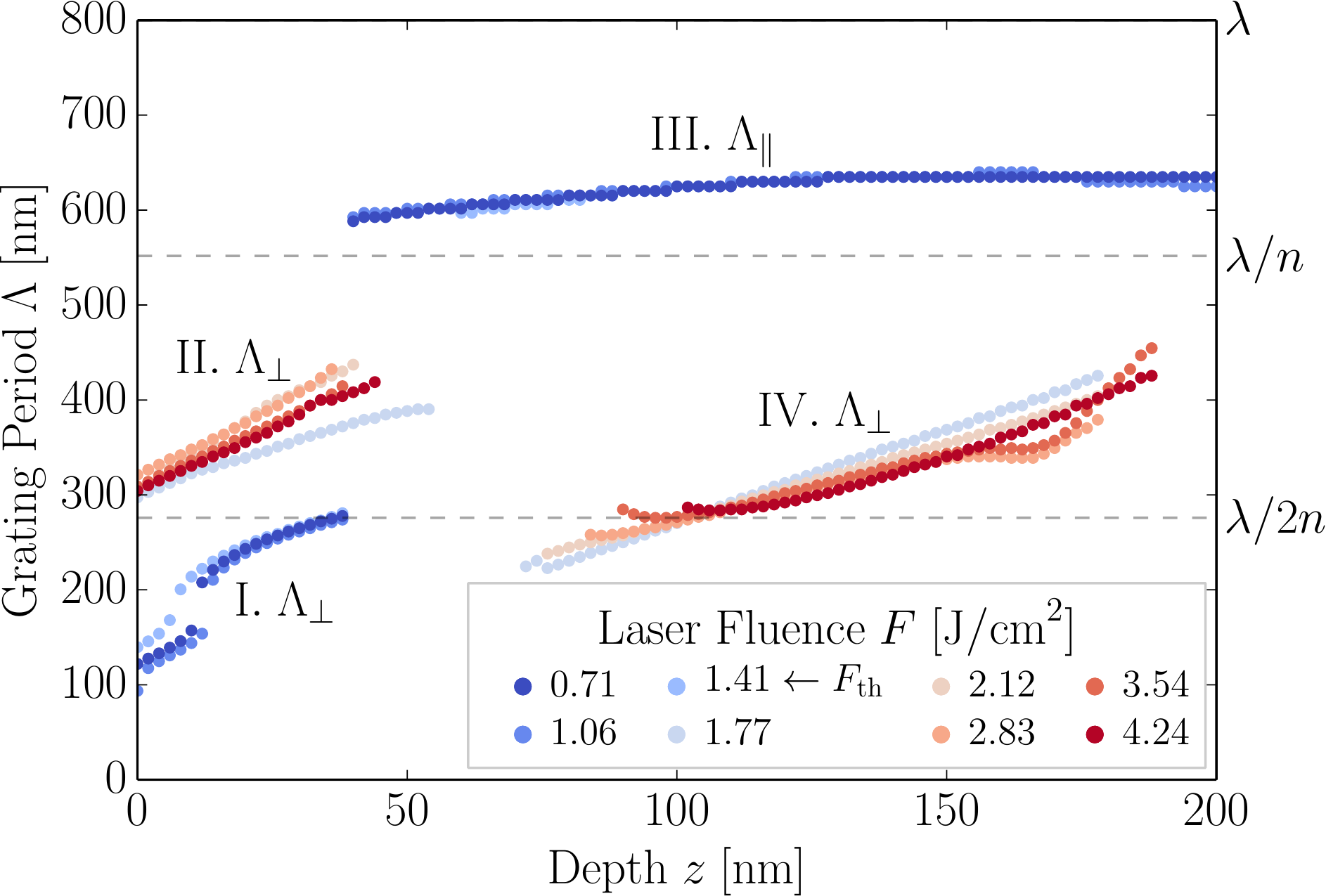}
\caption{(Color online) The grating period $\Lambda$ versus depth for different values of fluence.}\label{figLambda}
\end{figure}

\begin{figure}
\centering
\hspace{-5mm}
\includegraphics[width=0.96\columnwidth]{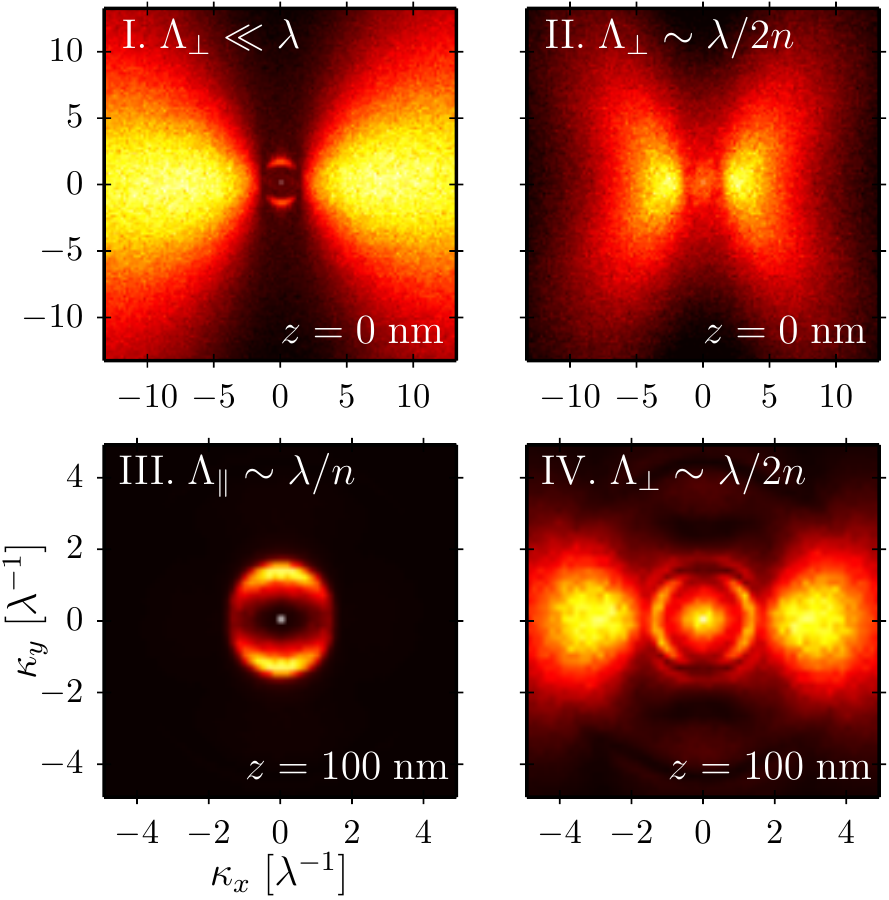}
\caption{(Color online) Averaged Fourier transforms of the four distinct structures identified in Fig. \ref{figLambda}. I and III are representative of below threshold results and are coherent with the Sipe theory. II and IV are representative of above threshold results, after the self-reconfiguration of the plasma patterns.}\label{figFourier}
\end{figure}

We investigated the dependence of the plasma patterns periods with respect to the laser fluence. To precisely measure $\Lambda_\perp$ and $\Lambda_\parallel$, we first reduce the noise in the Fourier transforms by averaging over 100 simulations, each with a reshuffled surface roughness, for several values of laser fluence. We then use the maximal values along the $\kappa_x$ and $\kappa_y$ axes to calculate the dominant periods $\Lambda_\parallel=\lambda/(\kappa_{y}|_\mathrm{max})$ or $\Lambda_\perp=\lambda/(\kappa_{x}|_\mathrm{max})$. The results are displayed in Fig. \ref{figLambda} in which four qualitatively distinct patterns are identified. Averaged Fourier transforms of each distinct structure are displayed in Fig. \ref{figFourier}.

From our numerical analysis, it is clear that the nature of the final plasma patterns qualitatively changes around a fluence threshold $F_\mathrm{th}\sim 1.41$ J/cm$^2$. Below threshold, where feedback is weak and static mode analysis applies, we observe two distinct patterns (see Fig. \ref{figLambda}). For $z<50$ nm, we find structures with $\Lambda_\perp \ll \lambda$ (I) that fit the near-field interference patterns \cite{zhang2015}, also present in the Sipe theory \cite{skolski2012}. Deeper in the sample, for $z>50$ nm, structures with $\Lambda_\parallel\sim\lambda/n$ (III) fit the far-field interference patterns in transparent media. Spectral signature of these patterns is found in Fig. \ref{figFourier}.

Above threshold, both near-field and far-field patterns shift towards a preferred periodicity of $\Lambda_\perp \sim \lambda/2n$ (II and IV). Since the plasma pattern self-reconfiguration is then dominated by transverse reflections, the natural explanation for this periodicity shift is the formation of standing waves along the $x$-axis \cite{buividas2011}. Light deflected towards $+x$ and $-x$ interfere to form a standing wave with antinodes separated by half wavelengths $\lambda/2n$ at every maxima of the plasma density. The condition for positive feedback is then met at this periodicity. It is important to note that the standing wave explanation is only valid for single or few-pulse expositions. In the multiple pulse regime, interpulse feedback mechanisms come into play and have to be included explicitly.

\section{Conclusions}\label{sec:conclusions}
In conclusion, we have simulated the growth of laser-induced plasma in the presence of dynamical intrapulse feedback, absent from the current LIPSS theory. We have demonstrated how the local and dynamic nature of the interactions between plasma and light in a transparent medium is responsible for an ultrafast femtosecond self-reconfiguration of the near-surface periodic patterns. Our analysis suggests that a 3D dynamical treatment of the plasma density is necessary to properly describe laser-induced surface patterns on dielectrics and the intrapulse feedback. We have provided a plausible mechanism, based on a simple stability analysis of the local feedback effects, that explains both the orientation and the periodicity of the laser-induced plasma patterns. Finally, we have demonstrated that self-organization of plasma structures, in agreement with the characteristics of laser-induced volume nanogratings, is possible without volume defects. We have shown that transverse mode coupling is enough to initiate the growth of the self-organized plasma structures when transverse reflections dominate over longitudinal ones. 

\subsection*{Acknowledments}
The authors acknowledge the financial support from the Natural Sciences and Engineering Research Council of Canada (NSERC) through the Canada Excellence Research Chair in Photonics Innovations. Computations were made on the supercomputer Guillimin from McGill University, managed by Calcul Qu\'ebec and Compute Canada. The operation of this supercomputer is funded by the Canada Foundation for Innovation (CFI), the Minist\`ere de l'\'Economie, de la Science et de l'Innovation du Qu\'ebec (MESI) and the Fonds de recherche du Qu\'ebec - Nature et technologies (FRQ-NT). Finally, the authors thank the EPOCH development and support team for their precious help. EPOCH development was funded by the UK EPSRC grants EP/G054950/1, EP/G056803/1, EP/G055165/1 and EP/ M022463/1.

\end{document}